\documentclass{aa}  
\usepackage{graphicx}

\usepackage{txfonts}

\def\0{\phantom0}

\def\zl{z$_{\rm lens}$}
\def\kms{km s$^{-1}$}

\begin{document}
\newcommand{\obj}{SDSS J0924+0219}

   \title{COSMOGRAIL: the  COSmological MOnitoring of \\ \vspace*{1mm}
    GRAvItational Lenses  II.  \thanks{Based on observations made with
    the ESO-VLT Unit Telescope 2 Kueyen (Cerro Paranal, Chile; Program
    074.A-0563, PI: G.  Meylan) and on data obtained with the NASA/ESA
    Hubble Space Telescope (Program  HST-GO-9744, PI: C.~S.  Kochanek)
    and   extracted  from the data  archives   at  the Space Telescope
    Science Institute,  which   is  operated by  the    Association of
    Universities for Research in  Astronomy, Inc., under NASA contract
    NAS~5-26555.}}
   
   \subtitle{SDSS J0924+0219: the  redshift of the lensing galaxy,  \\
    the quasar spectral variability and the Einstein rings}
   
   \titlerunning{COSMOGRAIL~II: SDSS J0924+0219: lens redshift, 
    quasar source spectra, and Einstein rings}

   \author{A. Eigenbrod\inst{1} \and F. Courbin\inst{1} \and S. Dye\inst{2}
        \and G. Meylan\inst{1} \and D. Sluse\inst{1} \and C. Vuissoz\inst{1}
        \and P. Magain\inst{3}}


    \institute{
     Laboratoire d'Astrophysique, Ecole Polytechnique F\'ed\'erale
     de Lausanne (EPFL), Observatoire, CH-1290 Sauverny, Switzerland
     \and
     School of Physics and Astronomy, Cardiff University,
     5 The Parade, Cardiff, CF24 3YB, UK
     \and 
     Institut   d'Astrophysique et  de  G\'eophysique, Universit\'e de
     Li\`ege, All\'ee du 6 ao\^ut 17, Sart-Tilman, Bat B5C, B-4000 Li\`ege,
     Belgium}

   \date{Received ... ; accepted ...}

 
  \abstract 
  {} 
  {To  provide the  observational    constraints required to  use  the
  gravitationally lensed quasar \obj\ for the determination of H$_0$ from
  the time   delay method.  We measure here the
  redshift of the lensing galaxy,  we show the spectral variability of
  the source, and we resolve the lensed host galaxy of the source.}
   {We present  our  VLT/FORS1 deep spectroscopic observations  of the
    lensed quasar \obj, as well as archival HST/NICMOS
    and  ACS images  of  the   same  object.  The  two-epoch  spectra,
    obtained  in the Multi Object   Spectroscopy (MOS) mode, allow for
    very accurate flux calibration   and spatial deconvolution.   This
    strategy provides  spectra for  the  lensing
    galaxy and for the quasar images A and B, free of any mutual light
    contamination.   We deconvolve the  HST  images as well,
    which reveal a double Einstein  ring. The mass distributions 
    in the lens, reconstructed in several ways, are compared.}
    {We determine the redshift  of the lensing galaxy  in \obj: \zl$ =
    0.394\pm0.001$.   Only slight spectral  variability is seen in the
    continuum of quasar images A and B, while  the C~III], Mg~II and Fe~II
    emission lines display obvious changes. The flux ratio between the 
    quasar images  A and B is the same  in the
    emission lines and  in the continuum.
    One of the Einstein rings found using deconvolution
    corresponds  to the  lensed quasar  host  galaxy at $z=1.524$ and  a
    second bluer one, is the image  either of a star-forming region in
    the host galaxy, or of another unrelated lower redshift object.  A
    broad range  of lens models give  a satisfactory fit to the data.
    However, they predict very different time  delays, making \obj\ an
    object  of particular  interest  for  photometric  monitoring.  In
    addition, the   lens  models reconstructed  using  exclusively the
    constraints  from  the Einstein rings,   or using  exclusively the
    astrometry  of the quasar    images, are  not compatible.     This
    suggests that multipole-like structures play  an important role in
    \obj.}
    {}
   \keywords{Gravitational  lensing:  quasar,  microlensing,   time
       delay --  Cosmology: cosmological parameters,  Hubble constant,
       dark matter.}

   \maketitle

\section{Introduction}

\begin{figure*}[t!]
\begin{center}
\includegraphics[width=13cm]{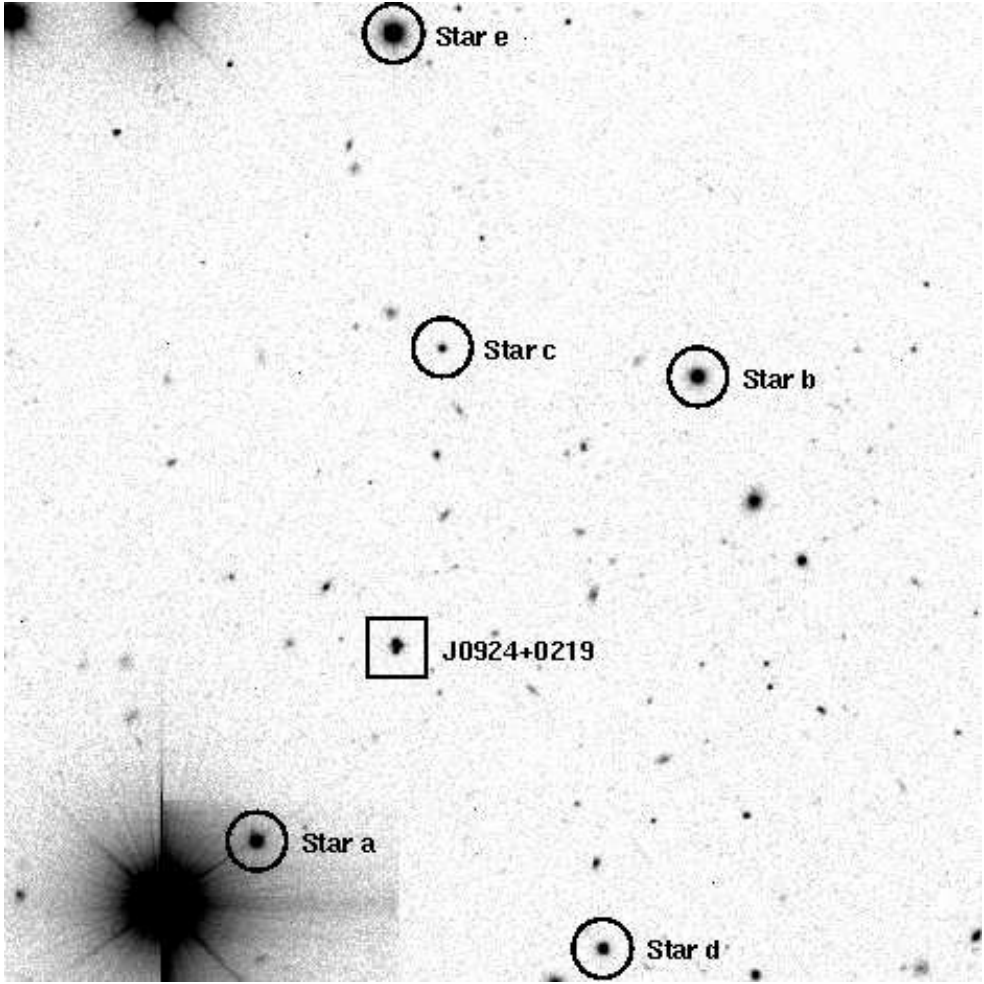}
\caption{$R$-band VLT image of \obj, where objects are labeled
  following Inada et al. (\cite{inada}). The stars a, c,  d, and e are
  used to compute the PSF spectrum (see text).   Only stars a, d and e
  are  used to derive the  relative flux  calibration between each MOS
  mask. The field of  view is $3.4  \arcmin \times 3.4 \arcmin$, North
  is to the top, East to the left.}
\label{field}
\end{center}
\end{figure*}

COSMOGRAIL   is a multi-site   optical  monitoring campaign  of lensed
quasars.  Following the  original work by  Refsdal (\cite{Refsdal64}),
its goal  is   to measure, with an    accuracy close  to  one  percent
(Eigenbrod  al.  \cite{eigenbrod}), the  so-called time  delay between
the images of most gravitationally  lensed quasars.  These time delays
are used in combination with  lens models and detailed observations of
individual systems to infer  the value of  the Hubble parameter H$_0$,
independent of any standard candle (e.g., reviews by Courbin et al.
\cite{courbin2002}, Kochanek \cite{koko_saasfee}).  

The present work is devoted to the quadruply imaged quasar
\obj\ (Inada et al.  \cite{inada}) at z = 1.524, discovered in the course of the
Sloan  Digital   Sky Survey   (SDSS).    This object   is particularly
interesting because of its anomalous image  flux ratios, the origin of
which is unclear.  It has been argued that the faintest image of \obj,
which is located at a saddle point of the  arrival-time surface, could
be demagnified either from star microlensing (Schechter et al.
\cite{Schech2004},  Keeton  et  al.   \cite{keeton2005})  or  subhalos
microlensing (Kochanek \& Dalal, \cite{koko2004}).

We analyse  here our deep optical  spectra of  \obj\ obtained with the
ESO Very Large Telescope (VLT).  These spectra are used to: 1- measure
the   redshift   of the  lensing  galaxy,   2-  estimate  the spectral
variability of the quasar, 3- measure  the flux ratio between images A
and B of \obj, in the continuum and  the broad emission lines.  Hubble
Space Telescope (HST) ACS  and NICMOS images  from the  STScI archives
are    deconvolved   using   the MCS     algorithm     (Magain et  al.
\cite{magain98}) which unveils two  Einstein  rings. One of the  rings
corresponds to  the host galaxy  of the quasar  source and is  used to
constrain the  lens  models.  The second  one   is probably  due to  a
star-forming region  in the  host galaxy of  the  quasar source  or to
another unrelated object.

\section{VLT Spectroscopy}

\subsection{Observations}

Our spectroscopic  observations of \obj\ are  part of a low dispersion
spectroscopic survey aimed  at measuring  all unknown lens  redshifts.
They are     acquired  with the FOcal     Reducer  and low  dispersion
Spectrograph (FORS1), mounted on the ESO Very Large Telescope, used the
MOS   mode (Multi  Object    Spectroscopy)  and the  high   resolution
collimator.  This configuration allows the simultaneous observation of
a  total   of     8     objects over      a    field  of      view  of
$3.4\arcmin\times3.4\arcmin$ with   a  pixel   scale  of  $0.1\arcsec$
(Fig.~\ref{field}).  The G300V   grism, used in combination  with  the
GG435 order sorting filter, leads to the useful wavelength range 4450
\AA\ $ <\lambda< $ 8650  \AA\ and to a  scale of $2.69$ \AA\ per pixel
in the spectral  direction.    This setup has  a  spectral  resolution
$R=\lambda/\Delta  \lambda  \simeq  300$   at the   central wavelength
$\lambda=5900$ \AA, which   translates  in velocity space  to  $\Delta
v=\textrm{c} \Delta \lambda / \lambda \simeq 1000$ \kms.

\begin{figure}[t!]
\begin{center}
\includegraphics[width=8.8cm]{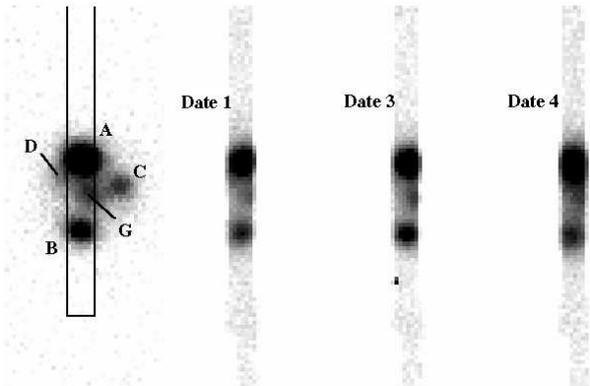}
\caption{$R$-band images of \obj. A short 30 sec
  exposure is shown on the left, where the quasar images A, B, C and D
  as well as  the   lensing galaxy,  are  indicated.   The  seeing  is
  $0.37\arcsec$ in  this image and the  pixel  scale is $0.10\arcsec$.
  The position of the $0.70\arcsec$ slitlets  is also indicated.  They
  correspond to three   epochs of  observations  with  very  different
  seeings (see  Table~\ref{refer}).   The slit has  not  moved  at all
  between the exposures, even when taken 15 days apart.}
\label{slits}
\end{center}
\end{figure}

\begin{table}[t!]
\caption[]{Journal of the VLT spectroscopic observations of \obj. The
seeing is measured on the spectrum of the PSF stars.
The exposure time is 1400 s for each of the 6 epochs.}
\label{refer}
\begin{flushleft}
\begin{tabular}{cccccc}
\hline
\hline
ID & Date & Seeing $[\arcsec]$ & Airmass & Weather \\
\hline
1 & 14/01/2005 & 0.66 & 1.188 & Photometric\\
2 & 14/01/2005 & 0.59 & 1.150 & Photometric\\
3 & 14/01/2005 & 0.46 & 1.124 & Light clouds\\
4 & 01/02/2005 & 0.83 & 1.181 & Photometric\\
5 & 01/02/2005 & 0.97 & 1.146 & Light clouds\\
6 & 01/02/2005 & 0.84 & 1.126 & Light clouds\\
\hline
\end{tabular}
\end{flushleft}
\end{table}

The slitlets of the MOS mask  are all 19\arcsec\ long and $0.7\arcsec$
wide,  which both avoids lateral  contamination  by the quasar image C
and matches   well the seeing values during    the observations.  Four
slits were centered on the foreground stars a, c,  d, e, while a fifth
slit is centered on images A and B of \obj, after rotation of the mask
to a suitable Position Angle (PA) (Fig.  \ref{slits}).  The spectra of
the stars are used both to compute the reference Point Spread Function
(PSF) needed for  the deconvolution and to  carry out a  very accurate
relative flux calibration. ``Through-slit'' images acquired just
before exposures  \#   1, \# 3,   \#  4 in  order  to  check the  mask
alignment are displayed in Fig.~\ref{slits}.

\subsection{Reduction and Deconvolution}

The   spectra  are      bias  subtracted    and   flat-fielded   using
IRAF\footnote{IRAF  is distributed by   the National Optical Astronomy
Observatories,  which are operated  by the Association of Universities
for Research in Astronomy, Inc.,  under cooperative agreement with the
National  Science Foundation.}.  The flat fields  for each slitlet are
created from 5  dome exposures, using  cosmic ray rejection.  They are
normalized   by  averaging 60 lines     along  the spatial  direction,
rejecting the 20 highest and 20  lowest pixels, then block replicating
the result to match the physical size of the individual flat fields.

Wavelength calibration is obtained from numerous emission lines in the
spectrum of Helium-Argon lamps.   The wavelength solution is fitted in
two dimension  to each   slitlet of the    MOS mask.  The fit  uses  a
fifth-order  Chebyshev polynomial along the   spectral direction and a
third-order  Chebyshev  polynomial fit  along  the  spatial direction.
Each  spectrum is   interpolated following  this  fit,  using  a cubic
interpolation.   This procedure ensures that  the  sky lines are  well
aligned with the columns of the CCD after wavelength calibration.  The
wavelength solution with respect to the reference lines is found to be
very good, with an rms scatter better than $0.2$ \AA\ for all spectra.

The  sky  background is  then removed by   fitting  a
second-order  Chebyshev polynomial in the    spatial direction to the
areas of the spectrum that are not illuminated by the object.

Finally, we perform the cosmic ray removal as follows. First, we shift
the spectra in order to align them spatially (this shift is only a few
tenths  of a pixel).  Second, we  create a  combined spectrum for each
object from the 6 exposures, removing the 2 lower and 2 higher pixels,
after applying  appropriate   flux  scaling.  The  combined   spectrum
obtained  in that way  is cosmic ray cleaned and   used as a reference
template to clean the individual spectra. We always check that neither
the   variable  seeing,  nor  the   variability  of  the quasar causes
artificial loss of data pixels.
 
Even though the seeing on most spectra  is good, the lensing galaxy is
close enough to the brightest quasar images A  and B to be affected by
significant contamination from the wings of  the PSF. For this reason,
the spectral version of MCS deconvolution algorithm (Magain et al.
\cite{magain98}, Courbin et al.  \cite{courbin}) is used in order to
separate the spectrum  of the lensing galaxy  from the spectra  of the
quasar  images.   The  MCS algorithm   uses  the spatial   information
contained in the  spectrum of a reference PSF,  which is obtained from
the slitlets positioned  on  the four isolated stars   a, c, d, and  e
(Fig.  \ref{field}).  The final normalized PSF is a combination of the
four PSF    spectra.  The   six  individual spectra    are deconvolved
separately,   extracted,  flux calibrated    as  explained in  Section
\ref{section:flux} and combined.   The spectrum of the lensing  galaxy
is extracted from  the ``extended channel''  of the deconvolved  data,
while  the  spectra  of  the quasar   images are  extracted   from the
``point-source channel'' (see Courbin et al. \cite{courbin}).

\subsection{Flux Calibration}
\label{section:flux}

\begin{figure}[t!]
\begin{center}
\includegraphics[width=8.8cm]{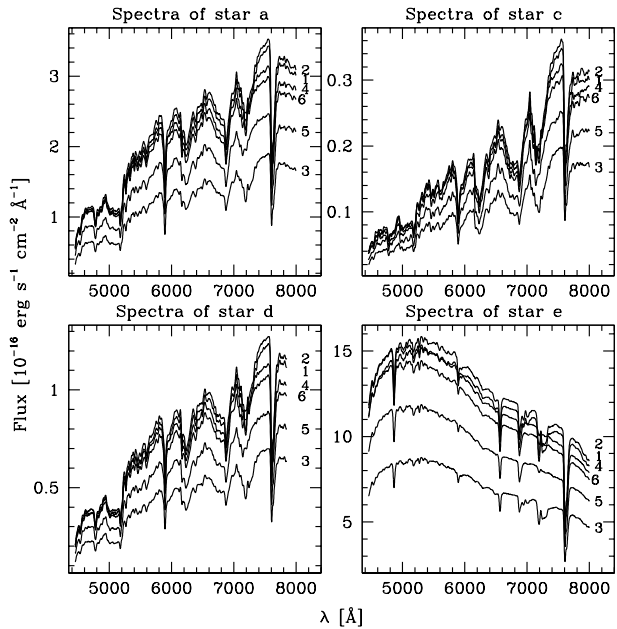}
\caption{The spectra of the foreground stars. The index on the right
  of each     spectrum indicates   the   exposure   number,  following
  Table~\ref{refer}.  Flux differences  are mainly due to the presence
  of light clouds on observation dates \# 3, \# 5 and \# 6.}
\label{PSF_spectra}
\end{center}
\end{figure}

Our absolute   flux  calibration is  based   on the  spectrum of   the
spectrophotometric standard  star Feige 66 taken on  the night of 2005
January 16.  The response function of the grism is determined for this
single epoch.  It is cross calibrated using stars observed in each MOS
mask in order to obtain a very accurate calibration across all epochs.
The spectra of  four stars  are  displayed in  Fig.~\ref{PSF_spectra},
without any deconvolution   and having used  a 4\arcsec\  aperture for
extraction. We  find significant differences in   flux between the six
epochs, that  need  to be corrected  for.  The  main causes  for these
differences are variable  seeing and variable  extinction due  to thin
cirrus during some  of  the   observations (Table~\ref{refer}).    The
effect of mask  misalignment  is excluded,   as can be  seen from  the
image-through-slit of Fig.~\ref{slits}.

\begin{figure}[t!]
\begin{center}
\includegraphics[width=8.8cm]{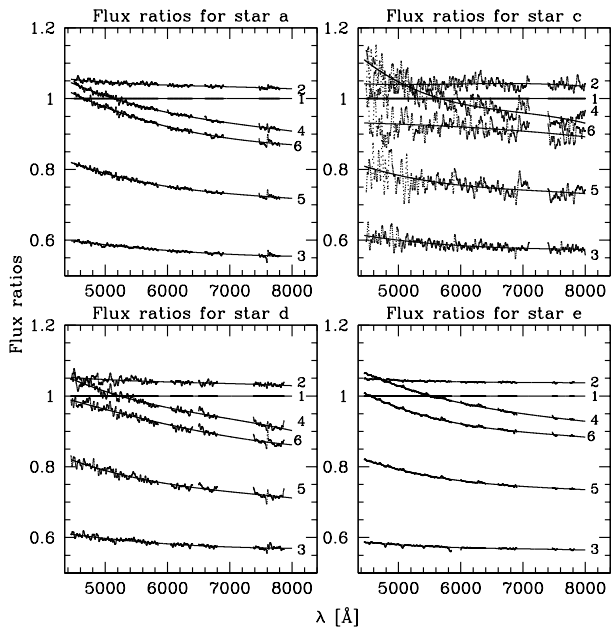}
\caption{Flux ratios between Date \# 1 and the 5 others, along with the
  third-order polynomial fits. We use the ratios of the 3 stars: a, d
  and e to determine the mean correction applied to  the quasar.  Star
  c, which  is much  fainter than  the  others, is excluded  from the
  final   calibration. The (small) parts   of  the spectra with strong
  atmospheric absorption are  masked during the  polynomial fit.  The
  peak-to-peak differences between the ratios computed using star a, d
  and e, differ by less than 2\%.}
\label{PSF_ratio}
\end{center}
\end{figure}

Assuming that the intrinsic flux of the foreground stars has not varied
between    the    six exposures,  and     taking    the  data \# 1   of
Table~\ref{refer} as a reference, we derive the flux ratio between
this reference epoch and  the six other  dates, for each star.   These
curves, fitted with a third-order polynomial, are  shown in Fig.
\ref{PSF_ratio}. The corrections computed in this way are found to be
very stable across the  mask:  the curves  obtained for two  different
stars  only  showed  slight   oscillations   with an amplitude   below
2\%.   This  is also   the  accuracy  of  the flux  correction between
different epochs.  A mean correction curve is then computed for each epoch
from all stars, except  star c which is much  fainter than the others,
and is applied to the  deconvolved spectra of  the quasars and of the
lensing galaxy.

\section{Extracted Spectra}

\subsection{The Lensing Galaxy}

\begin{figure}[t!]
\begin{center}
\includegraphics[width=8.8cm]{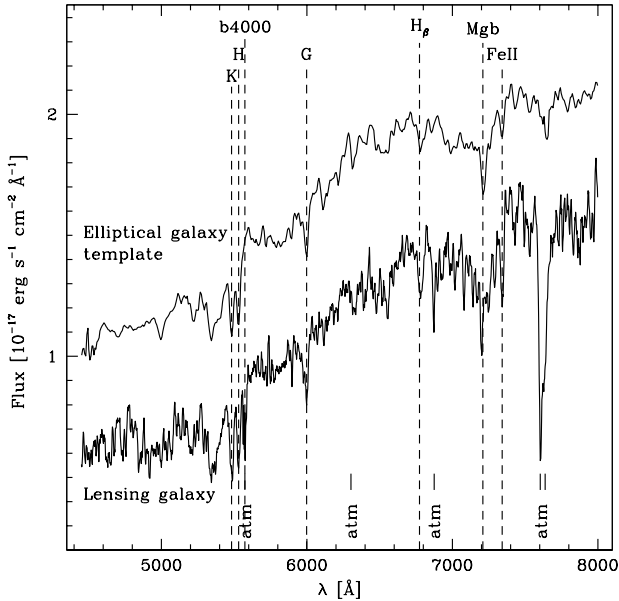}
\caption{Spectrum of the lensing galaxy in \obj, as obtained by
  combining the data for the 6 epochs, i.e.,  a total integration time
of 8400s.  The template spectrum of an elliptical galaxy at z=0.394 is
also shown for comparison  (Kinney  et al.  \cite{kinney}). All   main
stellar absorption lines  are well identified.  Prospects for a future
determination  of   the  galaxy's  velocity  dispersion  are therefore
excellent.}
\label{lens}
\end{center}
\end{figure}

The  six  deconvolved spectra  of  the  lensing  galaxy are extracted,
combined, and smoothed with a 5 \AA\ box (2 pixels).  Fig.  \ref{lens}
shows  the   final one-dimensional spectrum, where  the   Ca~II H \& K
absorption  lines are obvious, as well  as  the $4000$ \AA $\,$ break,
the G-band  typical for CH absorption,  the Mg band, and the H$\beta$,
and Fe~II absorption lines.  These features yield a mean redshift of
\zl$=0.394  \pm 0.001$, where  the  1-$\sigma$  error is the  standard
deviation  between all   the measurements  on   the  individual lines,
divided by the  square root of the number   of lines used.   We do not
consider the  $4000$   \AA $\,$ break   in  these calculations.   This
spectroscopic redshift falls very close to the photometric estimate of
$z=0.4$  by   Inada   et  al.   (\cite{inada}),    and  agrees with the
spectroscopic  redshift  of Ofek et  al.~(\cite{ofek05}).  In addition,
the absence  of emission lines  confirms a gas-poor early-type galaxy.
No trace of the quasar broad emission lines is seen in the spectrum of
the lensing galaxy, indicative of   an accurate decomposition of   the
data  into   the extended  (lens)  and  point  source (quasar  images)
channels.

\subsection{The Quasar Images}
 
\begin{figure}[t!]
\begin{center}
\includegraphics[width=8.8cm]{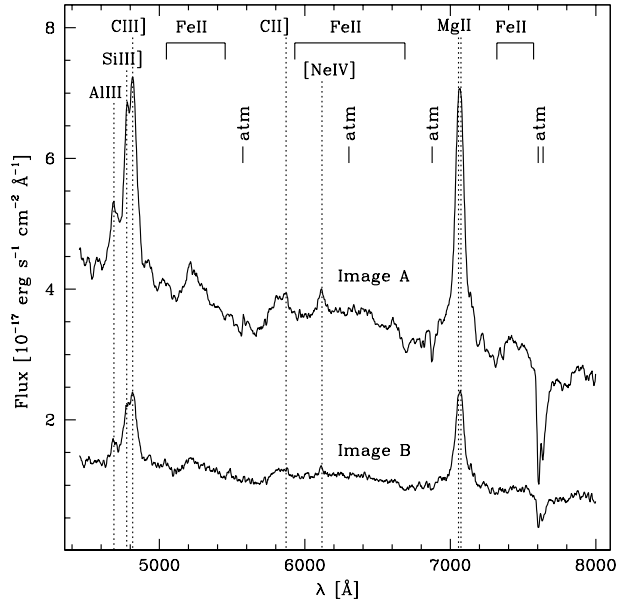}
\caption{Spectra of the quasar images A and B of \obj, as extracted
  from  the deconvolved  data. These figure  shows  the mean of  the 6
spectra taken for the  6 epochs, after  the flux calibration described
in Section~\ref{section:flux}.}
\label{quasars}
\end{center}
\end{figure}

The mean  spectra   of  quasar   images  A   and   B  are   shown   in
Fig.~\ref{quasars}, smoothed with  a $5$ \AA  $\,$ box.   The Al~III],
Si~III],  C~III], [Ne~IV] and  Mg~II  broad emission lines are clearly
identified.  A Gaussian fit to these 5  lines yield a mean redshift of
$1.524 \pm  0.001$ for image  A and $1.524  \pm 0.002$ for the fainter
image B.   The standard deviation between  the  fits to the individual
lines, divided  by the square  root  of the number  of  lines used, is
taken as the error bar.  These results are in excellent agreement with
the values obtained by  Inada et al.   (\cite{inada}), as well as  the
redshift from the SDSS database, who both report $z=1.524$.

\subsection{Variability of the Quasar Images}

\begin{figure}[t!]
\begin{center}
\includegraphics[width=8.8cm]{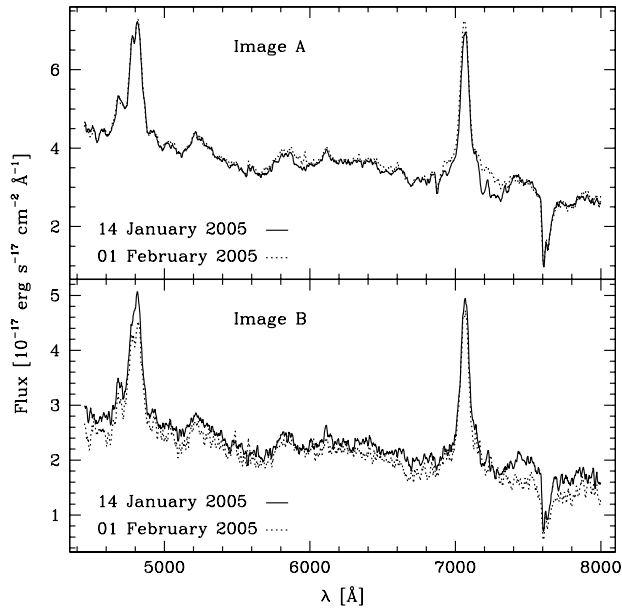}
\caption{The spectra of images A and B on 14 January and 1
February 2005 show a stable continuum for both images, but the
broad emission  lines  do vary  on  a time-scale   of two  weeks  (see
Fig.~.\ref{line_zoom}). }
\label{micro}
\end{center}
\end{figure}

The spectra of quasar images A and B are shown in Fig.~\ref{micro} for
2005  January 14 and  February  1.  These are the   mean of the  three
spectra obtained on each date,  smoothed with a  5 \AA\ box.  Although
the  continuum   shows little variation (only  B   has fadded slightly
between  our two  observing dates), there  are  obvious changes in the
broad emission lines of each quasar  image.  In image  A, the red wing
of the Mg~II  emission  line has brightened,  as   well as the   C~II]
emission line,  while in image  B,  the center  of the C~III] emission
line has become double peaked and  the Fe~II feature redwards of Mg~II
has  fadded.  A zoom on  these lines is shown in Fig.~\ref{line_zoom}.
The line  variations are   already visible   before  averaging  the  3
individual spectra  at a given  date and  in the not-so-blended quasar
images  of the raw  un-deconvolved  spectra.  We  can therefore safely
rule out any deconvolution artefacts due to PSF  variations in the MOS
mask.    In  addition, the  residual  images  after deconvolution (see
Courbin  et al.   2000    for  more details)  are   particularly good,
indicative of little or no PSF variations across the slitlet mask.

\subsection{Image flux ratio}

Keeton et al. (\cite{keeton2005}) have recently observed that the flux
ratio between the images of \obj\ is different in the continuum and in
the broad emission lines.  In their slitless HST/ACS observations, the
flux ratio  between A and  B is 2.60  in the emission lines, and about
3.5 in the continuum, i.e., the emission lines are 30\% different from
the continuum.

\begin{figure}[t!]
\begin{center}
\includegraphics[width=8.8cm]{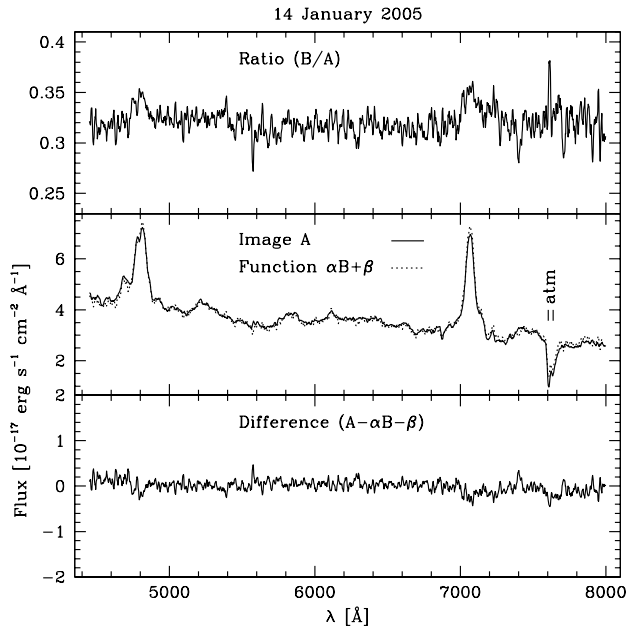}
\caption{Comparison between the spectra of images A and B taken on
  14  January    2005. The top panel    shows  the dimensionless ratio
  B/A. The mean ratio is $0.32$.  In the middle  panel, a first-order
  polynomial $\alpha$B$+\beta$ is fit to  the spectra of image A.  The
  best fit  is obtained with  $\alpha= 2.80 \pm 0.05$  and  $\beta =0.37$.  The
  difference in  flux   between A   and the  fitted  $\alpha$B$+\beta$
  polynomial is displayed  in the bottom  panel, and does not exceed a
  few percent of the flux.}
\label{BoverA_Jan}
\end{center}
\end{figure}

\begin{figure}[t!]
\begin{center}
\includegraphics[width=8.8cm]{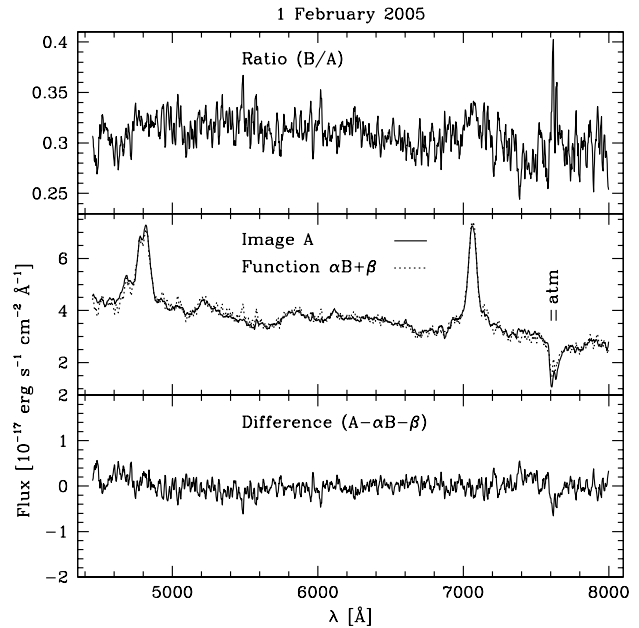}
\caption{Same as in Fig. \ref{BoverA_Jan} but for the spectra taken on
  1 February 2005. The  mean B/A ratio is $0.31$,  and the best fit of
  image A is obtained with $\alpha=2.86 \pm 0.05$ and $\beta =0.43$.}
\label{BoverA_Feb}
\end{center}
\end{figure}

We plot the flux ratio  between quasar image B and  A as a function of
wavelength at a given date (top panels in Figs.
\ref{BoverA_Jan} and  \ref{BoverA_Feb}).  This ratio  is close to flat,
with some small differences in the broad emission lines.

We  construct   the  spectrum $\alpha$B$+\beta$  and adjust   the
parameters  using a linear  least squares fit  so  that it matches the
spectrum of  quasar A.  The  result is shown in  the middle  panels of
Figs.~\ref{BoverA_Jan}  and \ref{BoverA_Feb}.   Almost  no trace of the
emission lines are seen in the difference spectra in the bottom panels
of  the figure.     Our  spectra   indicate  no strong    differential
amplification  of the continuum  and   broad  emission lines  in   the
components A  and  B of  \obj,  and the   small residual seen  in  the
emission lines in the bottom panels of Figs.~\ref{BoverA_Jan} and
\ref{BoverA_Feb} are an order of magnitude smaller than reported in
Keeton et al. (\cite{keeton2005}).

In the  15 days separating  the  observations, $\alpha$ has  changed by
only  2\%.   For  both   dates the residuals  of  the   fit are almost
perfectly flat,   indicating no   continuum  change. Only   asymmetric
changes in the emission lines are seen.

Finally, the flat flux ratio between image A and B shows that there is
no significant extinction by interstellar dust in the lensing galaxy.

\subsection{Intrinsic variations vs. microlensing}
\label{subsec:micro}

It is hard, with only two observing points, to infer the origin of the
spectral variations observed in  \obj.  Nevertheless, we see rapid (15
days)   and asymmetric changes  in  the  emission  lines of the quasar
images, and no strong  changes in the continuum.  Intrinsic variations
of quasars are usually stronger in  the continuum than in the emission
lines, and  they  are also longer  than the  two-week span we  observe
here.  Such rapid variations due to microlensing have  been seen in at
least   one  other    lensed  quasar:    HE~1104-1805  (Schechter   et
al.~\cite{Schechter03}).   \obj\   might be    a second   such   case.
Microlensing variability is   supported by the  photometric broad-band
data by  Kochanek et al.  (\cite{kokoIAU}),  showing that A and B have
very different light curves that are hard to match even after shifting
them  by the expected time delay.   However, microlensing usually acts
on the continuum rather than on  the emission lines of quasar spectra,
because  of the   much   smaller   size  of   the  continuum   region.
Differential amplification of  the continuum relative to the  emission
lines,  as observed by Keeton  et  al.~(\cite{keeton2005}), would be a
strong support to the microlensing hypothesis. Our spectra do not show
such  a  differential amplification, but we   note that our wavelength
range is very different from that of Keeton et al.~(\cite{keeton2005})
and that  they observed   in  May 2005,   i.e.,   3 months   after our
observations.

\begin{figure}[t!]
\begin{center}
\includegraphics[width=8.8cm]{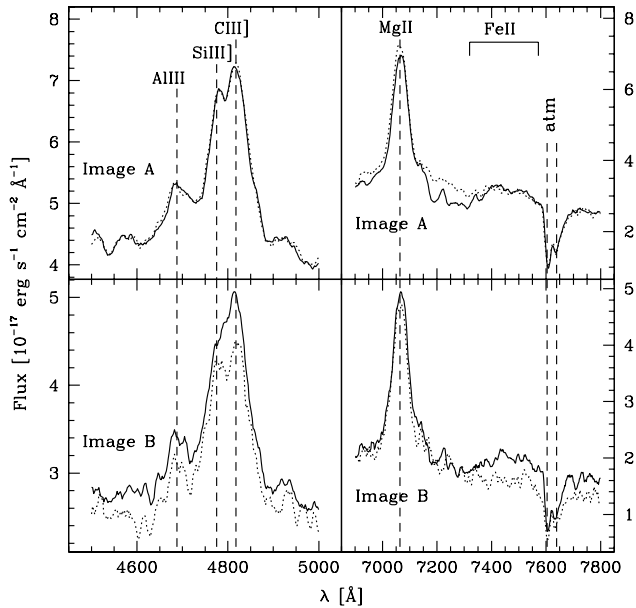}
\caption{Enlargements of Fig. \ref{micro} comparing the 
   broad emission lines of images A and B  on 14 January (solid curve)
  and 1 February 2005 (dotted curve).  Obvious  variations are seen in
  the red wing of the Mg~II in image A, in the center of the C~III] in
  image B. The Fe~II feature  redwards of Mg~II  in  image B has  also
  changed   by 20\%.  These  variations are asymmetric about the center
  of the lines. The asymmetry is different in C~III] and Mg~II.}
\label{line_zoom}
\end{center}
\end{figure}

Assuming microlensing  is the correct  interpretation of the data, its
strength depends upon  the   scale-size of the source,  with   smaller
sources  being   more    susceptible  to  large   magnification  (e.g.
Wambsganss  \& Paczynski  \cite{wambsganss}).  The  continuum emitting
region and the broad-line  region (BLR) of  a quasar can  appear small
enough  to undergo  significant  magnifications.   The limiting source
size   for  microlensing to  occur is   given  by the  Einstein radius
projected onto  the source plane.   This means that only structures in
the  source with sizes comparable to  or smaller than this radius will
experience  appreciable amplification.  The Einstein radius, projected
onto the source plane for microlenses with masses in the range $0.1
\,M_{\odot}< M < 10 \,M_{\odot}$ is 7\, $<R_E<$ 70\, light-days for a cosmology
with $\Omega_m=0.3$, $\Omega_{\Lambda}=0.7$ and h$_{100}$=0.65.

Kaspi  et al.  (\cite{kaspi}) derived sizes   for active galaxy nuclei
from reverberation mapping  of  the Balmer lines.   As a  function  of
intrinsic luminosity, they  found a global  scaling of  the broad-line
region (BLR) ranging from approximately $1$ to $300$ light days, which
compares  well  with the  Einstein  radius of the   microlenses in the
lensing galaxy of \obj.

The observations  also reveal  that the  broad  emission lines and the
continuum do not vary on the same time scale. Indeed, the continuum of
image    A  remains constant  over  the   15-day   time  span of  the
observations, while the broad emission lines vary.

Detailed microlensing simulations by Lewis \& Ibata (\cite{lewis})
show that the correlation between the magnification of the BLR and
the continuum source exists, but is weak.  Hence variations in the
broad emission lines need not be accompanied by variations in the
continuum.  This argument has been confirmed through observations of
other gravitationally lensed quasars (Chartas et al.  \cite{chartas},
Richards et al. \cite{richards}).

Another  observational fact   that  needs  some  enlightening is   the
asymmetric amplification  of    the    broad  emission   lines    (see
Fig. \ref{line_zoom}).  Such an amplification occurs for the C~II] and
Mg~II emission lines  in the  spectra of image   A.  The red  wings of
these lines are significantly more  amplified than the  blue ones.  An
explanation for  this is given  by Abajas  et al.  (\cite{abajas}) and
Lewis  \&  Ibata (\cite{lewis}), who show  that  emission lines can be
affected by substantial centroid  shifts and modification of  the line
profile.   Asymmetric  modification  of   the  line  profile   can  be
indicative of a rotating source.  Microlensing of  the part of the BLR
that  is rotating  away  from   us  would  then explain the   observed
asymmetric line amplifications.  This would  imply that a microlensing
caustic is passing at  the edge of the  broad line region, and  is far
enough from the continuum to leave it unaffected.

\section{HST Imaging}

Optical and near-IR images of \obj\ are available from the HST archive
in  the F555W,  F814W  and   F160W  filters.   The  F555W and    F814W
observations have been obtained on 18 November  2003 with the Advanced
Camera for Surveys (ACS)  and the Wide Field  Channel (WFC). The F555W
data consist of two dithered 1094 s exposures, each one being split in
two (CRSPLIT=2) in order to remove  cosmic rays.  Two consecutive 1148
s exposures have been taken through the F814W  filter, one hour later,
again   splitting  the exposure  time in   two.   Finally, the NICMOS2
observations,  taken on 2003   November  23,  consist  of 8   dithered
exposures, for a total  of 5312 s.  The   5-day period separating  the
optical and near-IR observations is of  the order of the expected time
delay between images A and B of the quasar.

\subsection{Image Deconvolution}
\label{deconv}

\begin{figure*}[t!]
\leavevmode
\begin{center}
\includegraphics[width=8.8cm]{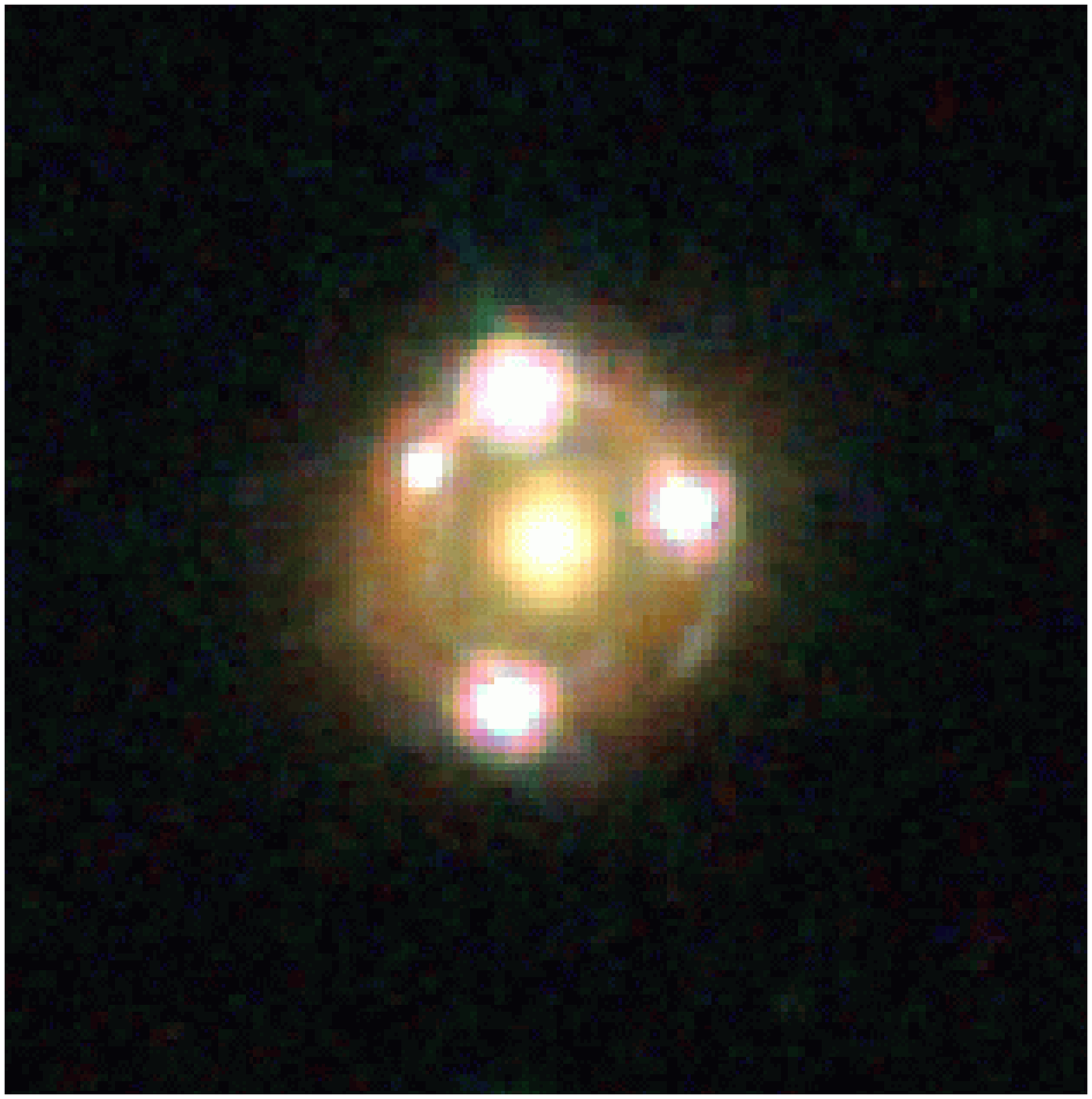}
\includegraphics[width=8.8cm]{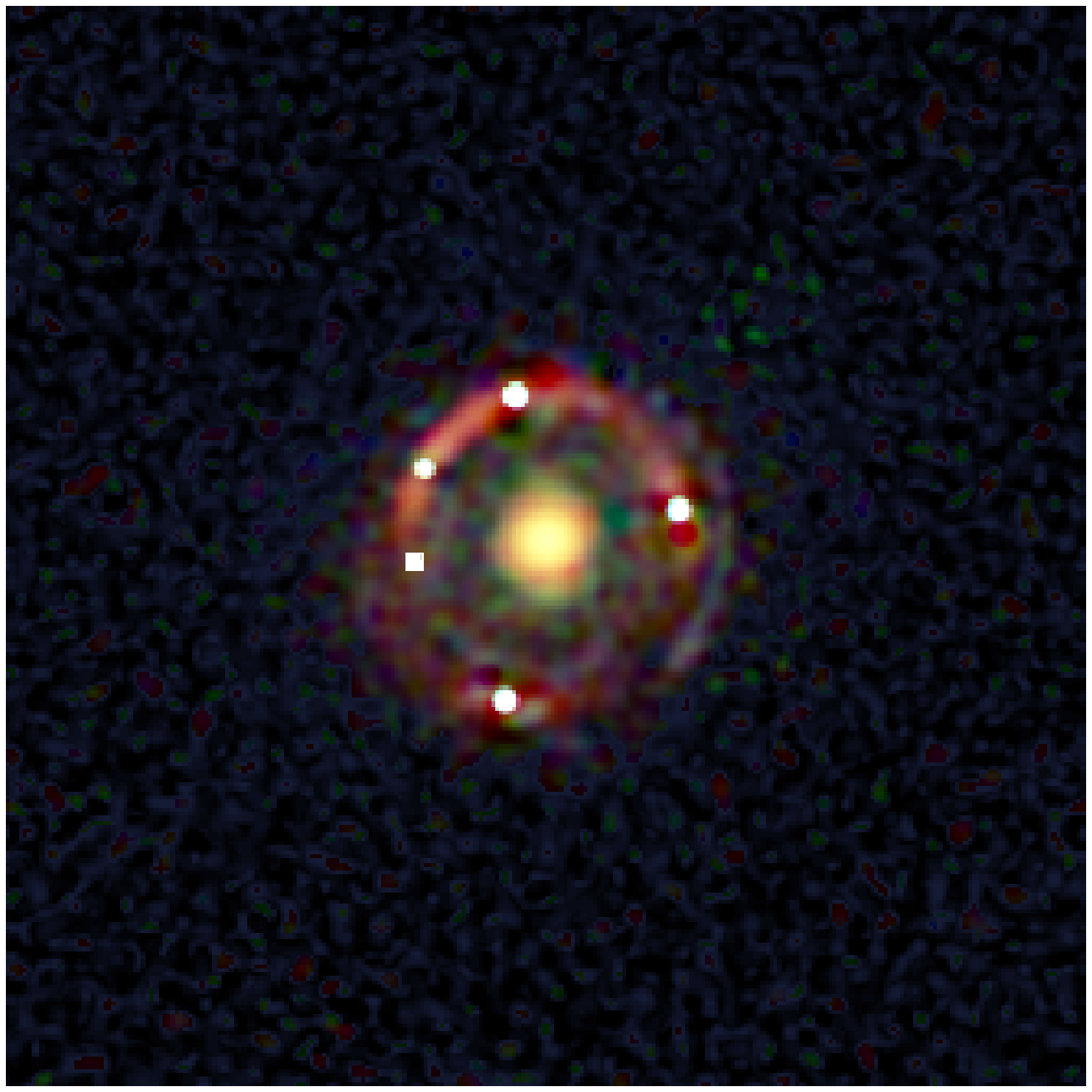}
\caption{{\it Left:} composite HST image using the observations
  through the  F555W,  F814W  and F160W   filters.  The  resolution is
  respectively 0.10\arcsec\ in  F555W   and F814W, and  0.15\arcsec\  in
  F160W. {\it Right:} deconvolved image.    It has  a pixel  size of
  0.025\arcsec\ and a resolution  of   0.05\arcsec.  The lensed   host
  galaxy of  the quasar is clearly seen  as red arcs
well centered on  the quasar images.  A  second set  of bluer
  arcs inside and outside the area delimited by the 
  red arcs is also revealed.  The field of
  view is 3.0\arcsec\ on a side. The image is slightly rotated relative to 
North, which is at PA=-2.67$^{\circ}$. East is to the left. 
The white square shows the position of the perturber found 
for the SIE and NFW models of Section~\ref{Simon}.}
\label{J0924_dec}
\end{center}
\end{figure*}

The  MCS algorithm (Magain   et    al.  \cite{magain98}) is   used  to
deconvolve all  images.    This algorithm  sharpens   the  images  and
preserves the flux of the original  data.  It also decomposes the data
into  a set  of analytical  point sources   (the quasar  images) and a
numerical ``extended channel'' which  contains all the features  other
than point sources,  i.e., the lensing  galaxy and  the Einstein ring.
All images are rebinned to a common pixel scale prior to deconvolution
and combined with  cosmic ray rejection.   The reference image adopted
to  carry out the  whole deconvolution work  is the  first image taken
through  the F814W filter,   i.e., image {\tt   j8oi33031} in  the HST
archive.  The position angle of this  reference image relative to the North is 
PA=-2.67$^{\circ}$.  All  the astrometry  in  the  following   is given  in the
coordinate  system of   this  image.   The  data   used  here  are the
pipeline-drizzled images available  from the archive.  The pixel scale
in the deconvolved  image is  half  that of the  original image, i.e.,
0.025\arcsec$\times$0.025\arcsec.   The spatial resolution is the same
in  all deconvolved images, i.e., 0.05\arcsec\ Full-Width-Half-Maximum
(FWHM).

As the HST PSF has significant spatial  variations across the field of
view, stars located  far away from  \obj\ on the plane  of the sky are
not  ideal for  use in the  image  deconvolution.  To  circumvent this
problem we have devised  an iterative procedure.  We first  deconvolve
the images with a fixed PSF, directly measured  from stars. This gives
a deconvolved image of the lens  and Einstein ring, that we reconvolve
with the PSF  and subtract from the  original  data.  A second PSF  is
re-computed from this new   lens- and ring-subtracted  image, directly
from the quasar images, following the procedure described in Magain et
al.~(\cite{magain05}).   This  is  similar  to a  blind-deconvolution,
where the  PSF is  modified  during the deconvolution  process.  A new
deconvolved image is created  with the improved PSF, as  well as a new
lens- and ring-subtracted  image.  We  repeat   4 times in  a row  the
procedure until  the  residual  map (Magain  et   al.~\cite{magain98},
Courbin   et al.\cite{courbin98}) is  flat and  in average  equal to 1
$\sigma$  after deconvolution,  i.e.,  until  the  deconvolved   image
becomes compatible with the data in the $\chi^2$ sense.

\begin{table}[t!]
\caption[]{Astrometry of \obj\ and flux ratio between the images.
All positions are given relative to the lensing galaxy in the 
coordinate system of our reference HST image {\tt   j8oi33031}. The
1-$\sigma$ error bar on the astrometry is 0.005\arcsec, mainly
dominated by the error on the position of the lensing galaxy. The
error  bar on the flux ratio is of the order of 10\% for images
B, C and 20\% for image D, and includes the systematic errors 
due to the presence of the Einstein ring (see text).}
\label{astrom}
\begin{flushleft}
\begin{tabular}{lccccc}
\hline\hline
Object    &      X        &  Y         & F555W & F814W & F160W  \\
          & (\arcsec)      & (\arcsec) &       &       &        \\
\hline
Lens      &  $+$0.000    & $+$0.000     &  $-$  &  $-$  &  $-$   \\
A         &  $-$0.185    & $+$0.859  & 1.00  & 1.00  & 1.00   \\
B         &  $-$0.246    & $-$0.948  & 0.51  & 0.46  & 0.44   \\
C         &  $+$0.782    & $+$0.178  & 0.39  & 0.34  & 0.32   \\
D         &  $-$0.727    & $+$0.430  & 0.06  & 0.06  & 0.03   \\
\hline
\end{tabular}
\end{flushleft}
\end{table}

\subsection{Results}

The deconvolved  images  through   the  three  filters  are  shown  in
Fig.~\ref{J0924_dec}, as a colour  composite image.  Two sets  of arcs
are clearly seen,  corresponding  to the host   galaxy of  the  source
quasar, and to   a bluer object  not centered   on the images   of the
quasar.  This arc is well  explained  by a  second lensed source  (see
Section~\ref{Simon}) which  is  either  a star-forming  region in  the
source, or another unrelated object.

Instead of using the   conventional version of the  MCS  deconvolution
algorithm, we use a version that involves a semi-analytical model for
the lensing galaxy. In this procedure, the analytical component of the
lensing galaxy  is either a two-dimensional  exponential disk, or a de
Vaucouleurs profile. All slight departures from these two profiles are
modeled in the form of a numerical array  of pixels which includes the
arcs as well. 

In all bands,  we  find that an  exponential disk  fits the data  much
better than  a  de Vaucouleurs   profile, which  is   surprising for an
elliptical   galaxy,    as     indicated   by   the     VLT   spectra.
Table~\ref{astrom} gives a summary of  our astrometry, relative to the
center of the fitted exponential disk.  The mean position angle of the
lensing galaxy, in the  orientation of the HST  image, is $PA =  -61.3
\pm   0.5^{\circ}$ (positive  angles  relative   to   the North  in  a
counter-clockwise   sense) and  the  mean ellipticity   is $e=0.12 \pm
0.01$,    where the  error  bars  are   the   dispersions  between the
measurements in the  three   filters.  We define the  ellipticity   as
$e=1-b/a$, where $a$ and  $b$ are the  semi-major and semi-minor  axis
respectively.   Note  that although the     formal error on  the  lens
ellipticity and PA   is small, the  data  show  evidence  for isophote
twisting.          The    effective radius     of      the   galaxy is
$R_e=0.50\pm0.05$\arcsec.

\begin{figure*}[t!]
\leavevmode
\begin{center}
\includegraphics[width=5.9cm]{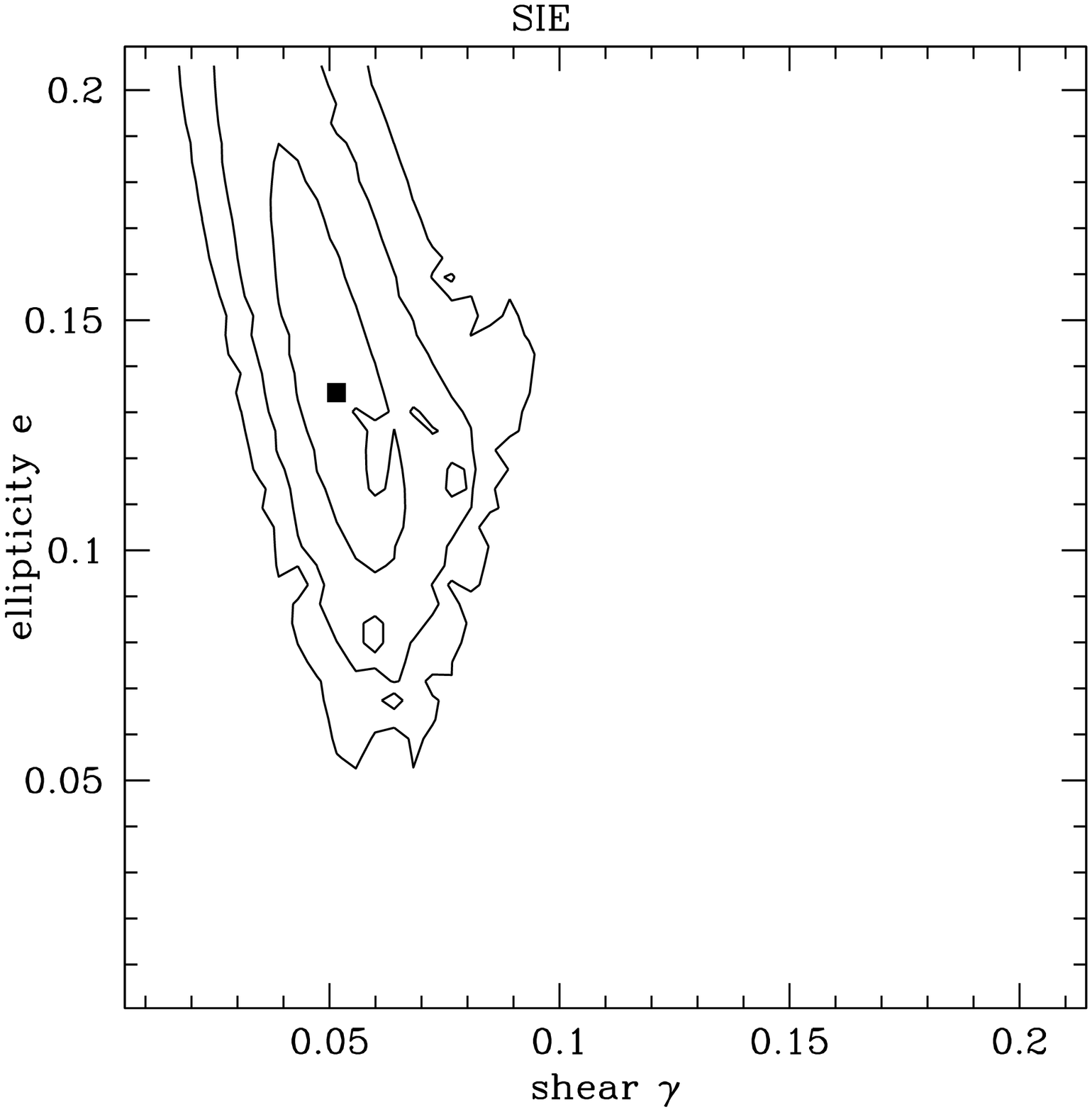}
\includegraphics[width=5.9cm]{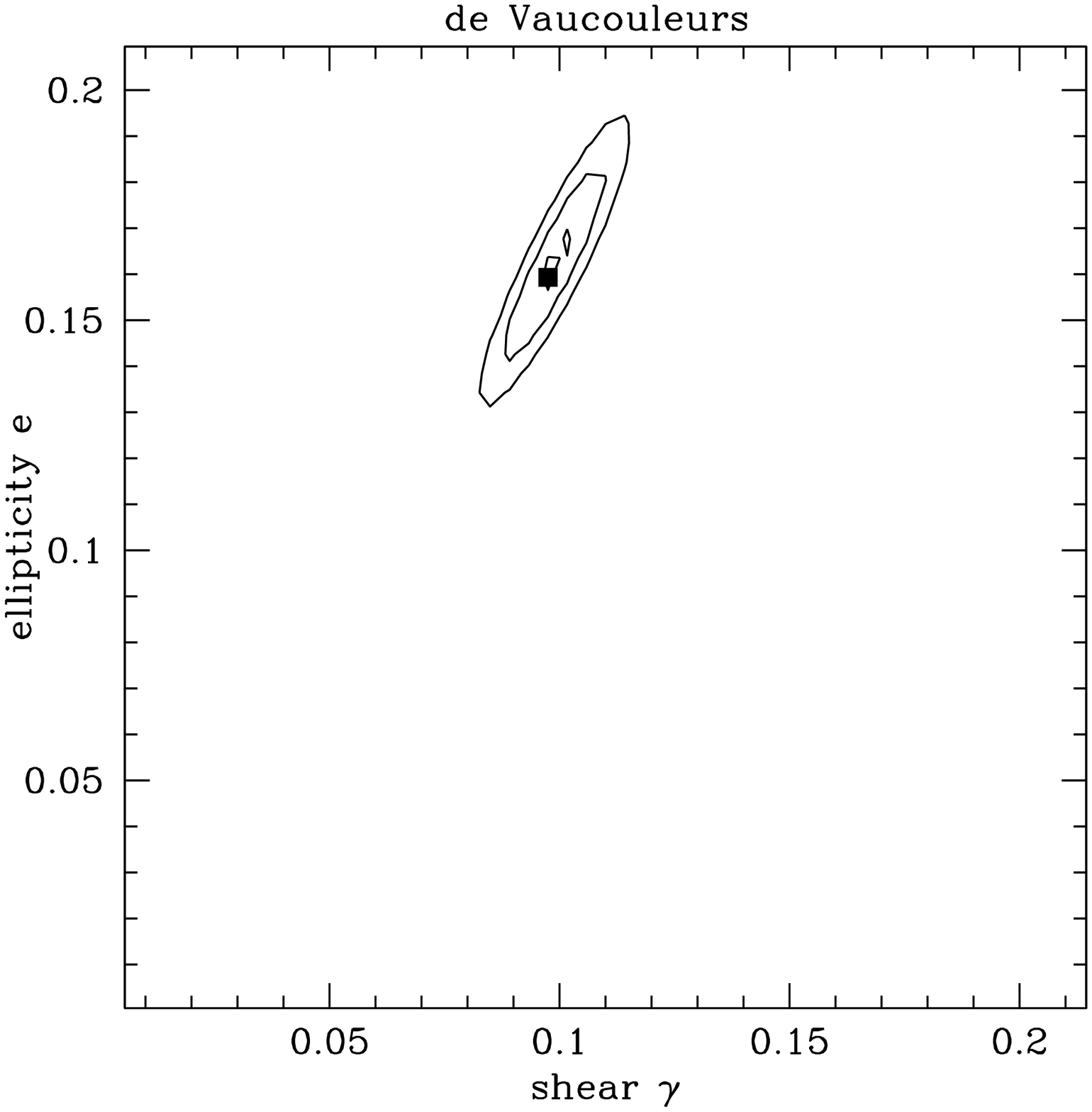}
\includegraphics[width=5.9cm]{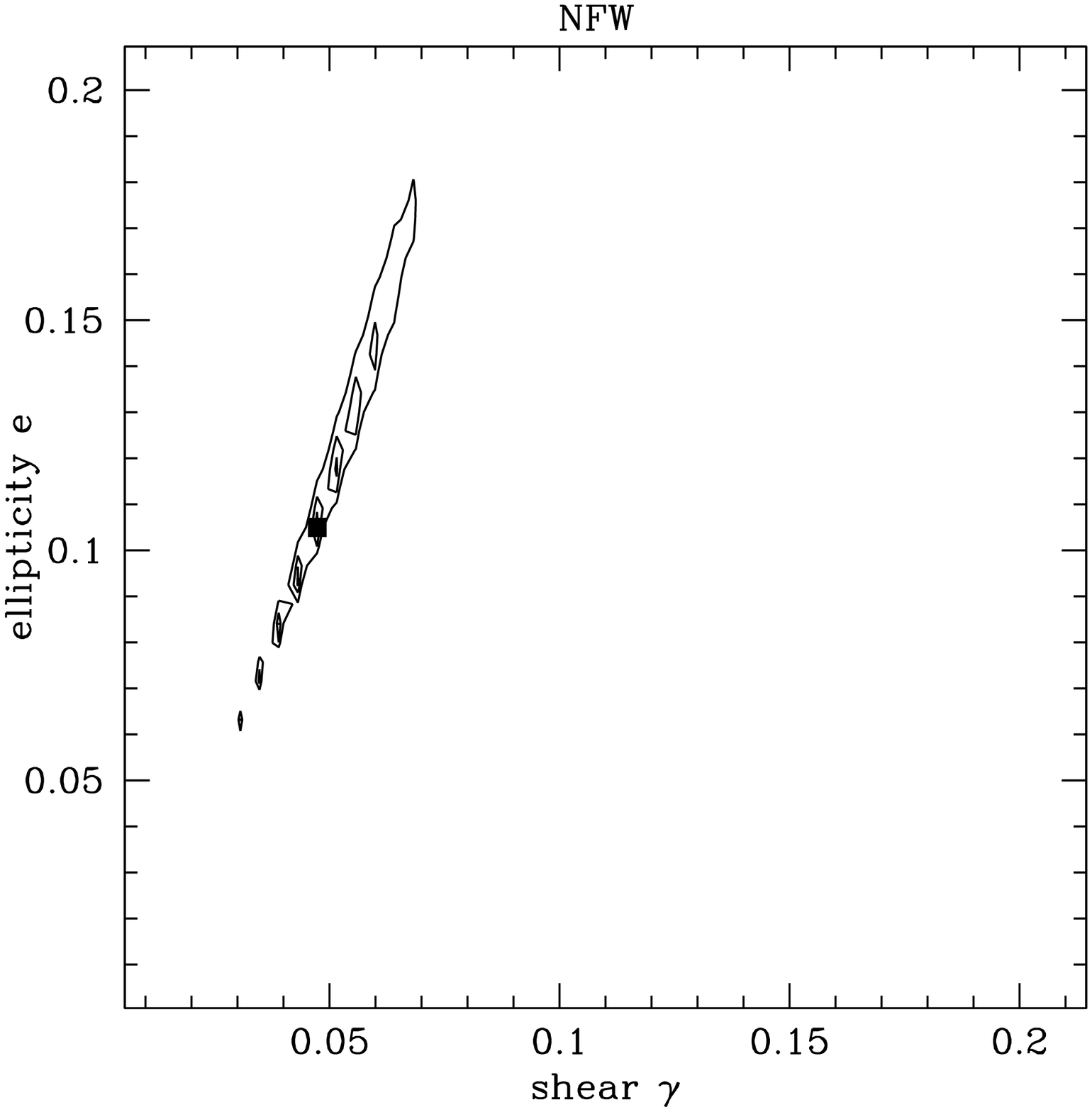}
\caption{The three plots give the reduced $\chi^2$ as a function of lens
ellipticity  $e$ and external  shear $\gamma$  for  the three analytic
models  used in the LENSMODEL package.   No constraint  is used on the
image  flux  ratios.   The  contours   correspond   to the   1, 2  and
3-$\sigma$, confidence  levels. The degeneracy between ellipticity and
shear is  clear. Only the NFW models  are (marginally) compatible with
no  external  shear at   all, as  also  suggested by   the semi-linear
inversion   of Section~\ref{Simon}. The   black  square in each  panel
indicated the   best  fit   model,   which  parameters are   given  in
Table~\ref{models}.}
\label{ellip_vs_shear}
\end{center}
\end{figure*}

\begin{table*}[t!]
\caption[]{Best-fit parametric models for \obj, obtained with the LENSMODEL 
package (Keeton~\cite{keeton_lensmodel}).   The position angles of the
lens $\theta_e$ and of the external  shear $\theta_{\gamma}$ are given
in degrees,  positive angles being counted  counter-clockwise relative
to the North.  The  coordinates $(x,y)$ of the  centres of the  models
are given in arcseconds, and the time delays  $\Delta t$ are expressed
in days relative to the leading image B. The extreme values 
for the time delays within the smallest 1-$\sigma$
region of Fig.~\ref{ellip_vs_shear} are also given.   
We adopt a $(\Omega_m, \Omega_\Lambda)=(0.3, 0.7)$ cosmology   
and h$_{100}$=0.65. All models have one degree of freedom.}
\label{models}
\begin{flushleft}
\begin{tabular}{llccccccccc}
\hline\hline
Model &  Parameters & $(x, y)$ & $e$ & $\theta_e$ & $\gamma$
& $\theta_{\gamma}$  & $\Delta t_{AB}$ & $\Delta t_{CB}$ & $\Delta t_{DB}$ & $\chi2$  \\
\hline
                   &                    & &  &  &  &  &   &  &  &   \\
SIE                & $b^{\prime}=0.87$ & $(-0.003,\, 0.002)$ & 0.13 & -73.1 & 0.053 & 65.4 & $5.7^{6.7}_{5.1}$ & $9.1^{10.4}_{8.2}$ & $6.2^{7.2}_{5.5}$ & 0.91  \\
                   &                    & &  &  &  &  &   &  &  &   \\
de Vaucouleurs     & $b=2.64$          & $(-0.004,\, 0.002)$ & 0.16 & -70.1 & 0.096 & 77.3 & $8.6^{8.9}_{8.1}$ & $13.8^{14.4}_{12.9}$ & $9.4^{9.7}_{8.8}$ & 1.41  \\
                   & $R_e=0.50$        & &  &  &  &  &   &  &  &   \\
                   &                    & &  &  &  &  &   &  &  &   \\
NFW                & $\kappa_s=0.70$   & $(-0.003,\, 0.001)$ & 0.10 & -72.0 & 0.047 & 65.4 & $4.9^{8.0}_{3.6}$ & $7.8^{12.7}_{5.8}$ & $5.4^{8.7}_{4.0}$ & 0.72  \\
                   & $r_s=1.10$        & &  &  &  &  &   &   &  &  \\
\hline
\end{tabular}
\end{flushleft}
\end{table*}

The flux ratios of the quasar images are derived from the deconvolved
images.  The MCS algorithm  provides the user  with the intensities of
all point sources in  the image, decontaminated  from the light of the
extended  features, such as the ring  in \obj\ and the lensing galaxy.
The error on  the quasar flux  ratio is dominated by the contamination
by  the Einstein  ring.    If  the intensity  of   a quasar  image  is
overestimated, this will create a ``hole'' in the deconvolved Einstein
ring at the quasar image position.  If it is underestimated, the local
$\chi^2$ at  the position of the quasar  image will become much larger
than 1 $\sigma$.  The flux  ratios in  Table~\ref{astrom} are taken  as
the ones giving at the same time a continuous Einstein ring without any
``hole'', and leading to a good $\chi^2$,  close to 1, at the position
of the quasar  images.  The  error  bars quoted in  Table~\ref{astrom}
are taken as  the   difference between these two   extreme solutions,
divided by 2.  They include both the random and systematic errors.

\section{Modeling}

Constraining the  mass distribution in \obj\  is not trivial. Firstly,
we do  not  have  access  to  the  true image  magnifications   due to
contamination by microlensing  and secondly, the light distribution of
the lensing galaxy is not very well  constrained.  The ellipticity and
position angle of the  lens change with surface brightness, indicative
of  isophote  twisting.  Measuring the  faintest  isophotes on the HST
data leads  to   PA $\simeq -25^{\circ}$, as is   adopted  by Keeton et
al.~(\cite{keeton2005})  in his models.   However, brighter  isophotes
and   fitting of a   PSF-deconvolved  exponential  disk  profile  yields
PA $= -61.3^{\circ}$.

As a blind test for the shape of the  mass distribution underlying the
light   distribution, and  without    using   any constraint on    the
ellipticity or PA  of  the lens,  we use the  non-parametric models of
Saha \& Williams  (\cite{saha04}).  Fitting  only the image  positions
returns  an asymmetric lens whose  major axis is aligned approximately
East-West (i.e., PA $= 90^{\circ}$).  Given the discrepancy between this
simple  model  and  the observed  light distribution,  we  test in the
following  a range of  models  with differing levels of  observational
constraints, in order to predict time delays.

\subsection{Parametric Models}

\subsubsection{Using the flux ratios}
\label{subsec:modA}

\begin{figure}[t!]
\begin{center}
\includegraphics[width=8.8cm]{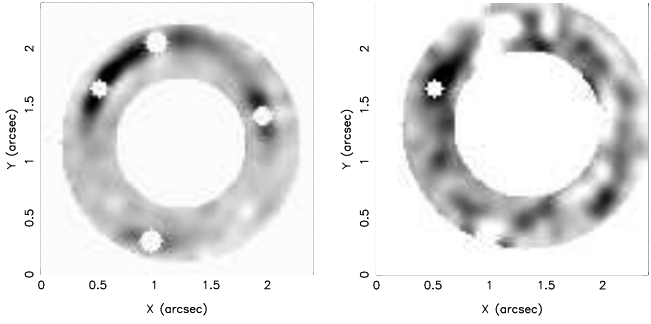}
\caption{Annular mask applied to the F160W (left) and F555W (right)
data with point sources masked out.  The annulus in the F555W image is
shifted  by 0.1\arcsec\ to  the left  and 0.2\arcsec\  to the top with
respect to the F160W image, to properly encompass the blue arc seen in
Fig.~\ref{J0924_dec}.}
\label{masked_rings}
\end{center}
\end{figure}

\begin{figure}[t!]
\begin{center}
\includegraphics[width=8.8cm]{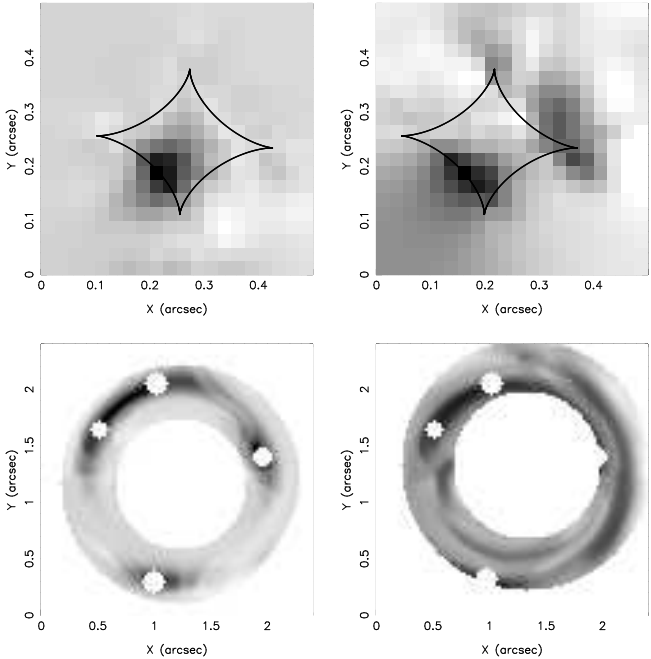}
\caption{Reconstructed source from F160W data (top left) and its
lensed   image (bottom left). A second   source lying on the rightmost
cusp caustic   (top right) is    reconstructed from  the  F555W  image
corresponding to the blue arc (bottom right).}
\label{recon_source}
\end{center}
\end{figure}

The LENSMODEL package (Keeton \cite{keeton_lensmodel}) is used to
carry out an analytical modeling of the lensing galaxy.  Three
lensing galaxy models are considered: the Singular Isothermal
Ellipsoid (SIE), the Navarro, Frenk \& White (\cite{nfw97}) profile
(NFW), and the de Vaucouleurs (\cite{devauc}) profile. In a first
attempt, we constrain these models with the lensing galaxy position,
the relative positions of the lensed images (Table~\ref{astrom}) and
their flux ratios (taken as the mean of the ratios measured in the
three F555W, F814W, F160W filters). If no external shear is included in
the models, we find a lens ellipticity of $e\simeq 0.3$ with a P.A.
$\theta_e \simeq 85^{\circ}$ and an associated $\chi^2 \simeq 200$.
The ellipticity and PA agree well with the models obtained
from the semi-linear inversion method of Warren \& Dye
(\cite{warren03}) (see Section~\ref{Simon}).

Next, we include external shear to the model.  The lens position angle
$\theta_e$, coordinates, and ellipticity agree better with the
measured values in the HST images.  The $\chi^2$ values remain bad
 ($\chi^2 \simeq 30$),
although  improved with respect to the models without external
shear.  The shear orientation is $\theta_{\gamma} \sim 60^{\circ}$
which is about in the direction of a bright galaxy located 9.5\arcsec\ 
away from \obj\ and at PA $= 53^{\circ}$.

The main contribution to the total $\chi^2$ is the anomalous
flux ratios between the images of \obj. In particular, the extreme flux
ratio between image A and image D of $\sim  15$, when these two images
are predicted to have approximately the same brightness. This is not
surprising because of the evidence of microlensing in image A
(Sect.~\ref{subsec:micro}) and of possible milli-lensing induced by massive
substructures. This lead us to the considerations presented in the 
next section.

\subsubsection{Discarding the flux ratios}

The modeling is similar to that of
Sect.~\ref{subsec:modA}. External shear is included but
the flux ratios are discarded. In order to use only models that have
one degrees of freedom (DOF), we have fixed the effective radius of
the de Vaucouleurs model to the observed value.  Given the number of
observational constraints, the NFW model would have zero DOF if all its
parameters were left free during the fit. We have therefore fixed
the orientation of the external shear in this model to the value we
found in the SIE+shear model. The best fit models are presented in
Table~\ref{models}, with (reduced) $\chi^2$ improved to values close
to 1.

We map the lens ellipticity vs.  external shear plane in order to
estimate the degree of degeneracy between these two parameters.  The
results are displayed in Fig.~\ref{ellip_vs_shear}. It is immediatly
seen that the 1-$\sigma$ ellipses of the different models only
marginally overlap. This is confirmed by the time delay values
summarized in Table~\ref{models} where we also give the extreme
values of the time delays within the 68\% confidence interval. The
minimum difference between the extreme time delays predicted with a
constant mass-to-light ratio galaxy (de Vaucouleurs) and by the more
physically plausible SIE model is about 8\%. Since the error
measurement on the time delay propagates linearly in the error budget,
even a rough estimate of the three time delays in \obj, with 8\%
accuracy will already allow to discriminate efficiently between flat
M/L models and SIE. Distinguishing between SIE and NFW is more
difficult as time delays predicted by NFW models differ by only 1\%
from the SIE time delays. Such an accuracy will be hard to reach in
\obj, that has short time delays and a short visibility period given
its equatorial position on the plane of the sky (see Eigenbrod et al.
2005).

\begin{table*}
\caption{Minimized lens model parameters and corresponding $\chi^2$.
Model  parameters are: $\kappa_0$  =  mass normalization in  arbitrary
units, $(x,y)$ = offset of lens model centre from lens optical axis in
arcseconds, $e$ =  ellipticity, $\gamma$ =  external shear, $\theta_e$
and $\theta_{\gamma}$ = PA in  degrees counted counter-clockwise  from
North.  In the  case of the NFW,  the  scale radius  is held fixed  at
$6''$  in  the minimization.  The  third column   gives the  number of
degrees of freedom (NDOF).  Subscript  '$b$'  refers to the  secondary
SIS in the dual component models (see text).}
\small
\begin{tabular}{l c c l}
\hline\hline
Model & $\chi^2_{min}$ & NDOF & Minimized parameters \\
\hline
SIE & 4280 & 3975 & $\kappa_0=100.0$, $(x,y)=(0.02, 0.04)$, $e=0.270$,
$\theta_e=86.0$  \\
NFW & 4011 & 3974 & $\kappa_0=100.0$, $(x,y)=(0.06, 0.06)$,
$e=0.187$, $\theta_e=84.9$ \\
Dual SIS & 4385 & 3974 & $\kappa_{0}=49.2$, $(x,y)=(0.00, 0.28)$,
$\kappa_{0b}=51.6$, $(x,y)_b=(-0.06, -0.33)$ \\
SIE$+$SIS & 4247 & 3972 & $\kappa_{0}=99.4$, $(x,y)=(0.04, 0.04)$,
$e=0.265$, $\theta_e=85.1$,  $\kappa_{0b}=2.1$, $(x,y)_b=(-0.79, -0.03)$ \\
NFW$+$SIS & 3971 & 3971 & $\kappa_{0}=98.0$, $(x,y)=(0.05, 0.08)$,
$e=0.206$, $\theta_e=83.1$,  $\kappa_{0b}=2.8$, $(x,y)_b=(-0.80, -0.09)$ \\
NFW$+ \, \gamma$ & 3992 & 3972 & $\kappa_{0}=100.0$, $(x,y)=(0.06, 0.06)$,
$e=0.168$, $\theta_e=86.0$, $\gamma=0.010$,  $\theta_{\gamma}=78.3$\\
\hline
\end{tabular}
\normalsize

\label{tab_sl_results}
\end{table*}

\subsection{Using the Arcs of the Lensed Sources}
\label{Simon}

The HST images of \obj\ reveal two sets of  arcs.  One is prominent in
the  near-IR (in red in  Fig.~\ref{J0924_dec}) and is well centered on
the quasar images. It is the lensed image of the quasar host galaxy. A
second set of bluer  arcs is best seen   in F555W. It is  off-centered
with respect  to the quasar images, indicating   either a companion to
the quasar host,  or an independent intervening  object along the line
of sight.

We  apply    the   semi-linear inversion  method   of    Warren \& Dye
(\cite{warren03}) to the arcs observed   in the F555W and F160W  data.
The method incorporates a linear matrix inversion to obtain the source
surface brightness  distribution  that  gives  the   best  fit to  the
observed lensed  image for a  given lens model.   This  linear step is
carried out  per trial lens  parametrisation  in a standard non-linear
search for the global best fit.

Dye \& Warren (\cite{dye05})  successfully apply this technique to the
Einstein  ring system 0047$-$2808.  They  demonstrate  that the  extra
constraints provided  by the image  of  the  ring results  in  smaller
errors on the  reconstructed lens model,  compared  to using  only the
centroids of the principal images as constraints in this system.

In  the  case of  0047$-$2808,  the  source is  a star  forming galaxy
without any point-like emission whereas the  image of \obj\ is clearly
dominated   by  the  QSO's  central  point  source.     To prevent the
reconstruction of \obj\ from being  dominated by the point source  and
because in    this section only the  reconstruction   of the  QSO host
emission is of interest, we masked out the four point source images in
the F555W  and F160W data supplied to  the semi-linear inversion code.
The    astrometry   of   the  quasar  images   is    not   used   as a
constraint. Fig. \ref{masked_rings} shows the masked ring images.

\subsubsection{Reconstruction results}

The   deconvolved F160W   and  F555W data   are  reconstructed with  6
different parametric lens  models.   Three of  these  are single  mass
component  models:    the singular isothermal    ellipsoid (SIE),  the
elliptical  NFW, and the   elliptical NFW  with  external  shear.  The
remaining three  test for asymmetry  in the lens  model by including a
secondary singular isothermal sphere (SIS) mass component that is also
free to move around in the lens plane and vary in normalization in the
minimization. These models are the dual SIS model, the SIE$+$SIS model
and the NFW$+$SIS model.

Since the F160W  data have the  highest signal to  noise arcs, we base
our lens modeling on these data and applied our overall best fit model
to  the F555W   data to  reconstruct the source.    In all  cases,  we
reconstruct   with   a  $0.5''\times0.5''$   source  plane  comprising
$10\times 10$ pixels. The reconstruction is not regularised, except in
Fig.  \ref{recon_source}  where first order regularisation (see Warren
\&  Dye \cite{warren03}) is  applied to  enhance  visualization of the
source.

Table~\ref{tab_sl_results} lists the   minimized parameters  for  each
model  and the  corresponding values of   $\chi^2$.  The SIE$+$SIS and
NFW$+$SIS models  clearly   fare better than  their   single component
counterparts, implying the  lens is asymmetric.   For the SIE$+$SIS, a
decrease  in $\chi^2$  of $\Delta \chi^2=33$   for 3 fewer degrees  of
freedom has a significance of 5.1 $\sigma$. The decrease of $\Delta
\chi^2=40$ for the NFW$+$SIS has a significance of 5.7 $\sigma$. Both
models consistently place the    secondary SIS mass component   around
$(-0.80'',-0.05'')$ with a normalization  of only $\sim 2.5$\% of  the
main component.

Interestingly,  the   elliptical       models    listed    in     Table
\ref{tab_sl_results} have ellipticities  close to those 
obtained  with  the  LENSMODEL  software, when   no external shear  is
considered.   When external shear  is  added to the  NFW model,  we do
indeed obtain  a significantly better fit  compared to the  NFW on its
own, but the results  differ from those listed in  Table~\ref{models}.
While     the  ellipticity    remains   almost    the    same  as   in
Table~\ref{models}, its PA differs by approximately 25$^o$.  Moreover,
we find  a  ten  times  smaller amplitude   for the  shear   using the
semi-linear inversion  than using LENSMODEL.   Note, however, that the
observed quasar image astrometry  is  used in the LENSMODEL  analysis,
whereas it is not in the present semi-linear inversion.  If we use the
lens model found by the semi-linear  inversion to predict the position
of the quasar images, we find poor agreement between the predicted and
the  measured  positions.  The global,  large  scale shape of the lens
found  by the  semi-linear  inversion is  well  adapted  to model  the
Einstein rings, which are very sensitive to azimuthal asymmetry in the
lens, but  additional smaller scale  structures are needed to slightly
modify  the  positions of the  quasar images  and make them compatible
with the measured  astrometry.  The disagreement  between the
astrometry  predicted    by LENSMODEL and the    one  predicted by the
semi-linear  inversion adds support  to the presence of multipole-type
substructures in the lens (e.g., Congdon \& Keeton~\cite{cong2005}).

The top left  plot in Fig.~\ref{recon_source} shows the  reconstructed
source corresponding to  the best fit  NFW$+$SIS  model for the  F160W
data.  The observed arcs  are explained by  a single  QSO host galaxy.
Note that in  this figure,   purely   to aid visualization,  we   have
regularised the   solution and plotted the   surface brightness with a
pixel  scale half that used  in the  quantitative reconstruction.  The
bottom left  corner of Fig.~\ref{recon_source} shows  the image of the
reconstructed source lensed by the best fit NFW$+$SIS model.

We then take the best fit NFW$+$SIS model in  order to reconstruct the
F555W data shown on the right in  Fig.~\ref{masked_rings}.  Note that
the annular  mask is shifted slightly  compared to the  F160W data, to
properly   encompass  the blue  arc.   The  reconstructed   source and
corresponding  lensed  image are   shown on the   right hand  side  of
Fig.~\ref{recon_source}.

There  are two distinct sources now  visible.  The QSO host identified
previously has again been reconstructed.  This is because its dominant
image, the bright arc  in the top left  quadrant of the ring, is still
present in the F555W data. A second  source, more diffuse and lying on
the  rightmost cusp  caustic is also  visible.   This second source is
responsible for the blue arcs.

The redshift of the second source remains unknown.  It could be a star
forming  object/region  lying  $0.2\arcsec  \cdot  D_s  \simeq 1200\,
h_{100}^{-1}$ pc away from the  quasar, i.e., it would  be part of the
host galaxy.
It is, however, not excluded that this second source is at a  
different   redshift than the quasar,   e.g.
located between  the  quasar and the  lens,  as it  is bluer  than the
quasar host galaxy.  If the latter  is true, \obj\  might  be a unique
object  to break the  mass  sheet degeneracy.  Unfortunately, the lens
modeling alone, does not allow to infer a redshift estimate.

\subsection{Note on the different types of models}

The two methods used above  differ in several respects. LENSMODEL
has a limited number of free parameters but  uses only the constraints
on  the astrometry of the quasar  images. While a qualitative representation
of the lensed host  galaxy of the  quasar source can be attempted, the
method does not allow a genuine fitting of the Einstein rings assuming
a (simplified) shape for the quasar host.

The semi-linear inversion carries  out a direct reconstruction  of the
lensed source as a whole, where each pixel of the  HST image is a free
parameter. As the quasar  images largely dominate  the total flux  of
the source, they need  to  be masked before  the reconstruction.   For
this reason it is not possible with this method, at the present stage of 
its development, to constrain the lens
model using {\it simultaneously} the astrometry of the quasar 
images and the detailed shape of the Einstein rings. 

Although the two methods used in  the present work are fundamentally
different and although they use very different observational constraints, 
they agree on the necessity to bring extra mass near image D of \obj.
Smooth lenses like the ones implemented in LENSMODEL have
PAs that differ by 10$^{\circ}$ from the one measured in the HST images.
In the orientation of Fig.~\ref{J0924_dec}, the mass distribution found
by LENSMODEL is closer to horizontal (PA$=-90^{\circ}$) than the light distribution, hence
giving larger masses next to image D. In the semi linear inversion,
the optimal position found for the SIS perturber is also close to image 
D. 

Given the above discussion, the poor determination of the lens PA is a main
limitation to the interpretation of the time delays in \obj. 
An alternative route is to determine the dynamical rotation axis
of the lens, a challenge which is now within the reach of  integral field 
spectroscopy with large telescopes and adaptive optics.

\section{Conclusions}

We  have  spatially deconvolved  deep sharp  VLT/FORS1 MOS  spectra of
\obj,    and  measured   the    redshift   of  the   lensing   galaxy,
\zl $= 0.394\pm0.001$, from numerous stellar absorption lines.  The
spectrum beautifully matches  the elliptical galaxy template of Kinney
et al. (\cite{kinney}).

The flux ratio between image A and B  is $F_A/F_B =  2.80 \pm 0.05$ on
2005 January 14, and  $F_A/F_B = 2.86  \pm  0.05$ on 2005  February 1,
i.e., it has not changed between the two dates given the uncertainties
on the flux ratios (Table~\ref{refer}).  For  each date, this ratio is
mostly  the same in the  continuum and in  the broad emission lines of
the quasar images  A and B.   This may seem in contradiction
with   Keeton   et    al.~(\cite{keeton2005})   who   see differential
amplification   of  the  continuum  relative  to   the lines,  but our
observing dates and setup are very different from theirs.

While the continuum of  images A and  B has  not  changed in 15  days,
there are  obvious and asymmetric changes in  some of the quasar broad
emission lines.  Microlensing of both A and B is compatible with this,
although somewhat ad hoc  assumptions must be  done on the position of
the  microcaustics relative to the quasar,  as well as on the relative
sizes of the continuum and broad line regions.

Deep HST imaging reveals two sets of arcs.  One corresponds to the red
lensed  host  galaxy  of   the quasar  and   defined an  Einstein ring
connecting  the quasar images.   The    other, fainter  and  bluer  is
off-centered  with  respect  to the quasar    images.  It is  either a
star-forming region in  the   quasar source host  galaxy,   or another
intervening object.

The lens ellipticity and  PA measured in the   HST images are hard  to
reconcile with  simple models without  external shear.  The model fits
improve when external  shear is added, even   though the predicted  PA
differs from the measured one by approximately $25^{\circ}$.
 
Models  of Section~\ref{Simon},  involving an  additional small (SIS)
structure to  the main lens always place  it along the East-West axis,
about 0.8\arcsec\  to the  East  of the main   lens, i.e., towards the
demagnified image D.  In addition, the models reconstructed using only
the Einstein  rings  do not predict  the  correct astrometry  for  the
quasar images.   Einstein rings constrain the  overall, large scale of
the lens.  Small deviations from this  large scale shape are needed to
match the quasar images  astrometry.  The discrepancy between
the  models using  the  rings and  the   ones using the  quasar  image
positions  therefore adds support to   the presence of  multipole-like
substructures in the lens of \obj.

Finally, the  range of time  delays  predicted by the  different  lens
models is large and   is very sensitive   to the presence of  external
shear and to  the determination of  the main lens  ellipticity and PA.
The   time delay measurement and    the  lens modeling, combined  with
integral field spectroscopy of the lens in \obj\ might therefore prove
extremely  useful to  map  the  mass-to-light ratio   in the lens,  by
comparing the lensing and dynamical  masses, to the light distribution
infered from the HST images.

\begin{acknowledgements}
The authors  would  like  to  thanks Dr.  Steve   Warren  for useful
discussions and the ESO staff at Paranal for the care taken with the
crucial   slit  alignment  necessary	to carry   out   the spectra
deconvolutions.   The HST archive  data  used  in this article  were
obtained in the  framework of the  CfA-Arizona Space  Telescope LEns
Survey (CASTLES, HST-GO-9744,  PI: C.S.  Kochanek).  PM acknowledges
support from the PSS  Science Policy (Belgium)  and by PRODEX (ESA).
COSMOGRAIL is financially   supported by the Swiss  National Science
Foundation (SNSF).
\end{acknowledgements}

\end{document}